\documentclass[namedreferences]{SolarPhysics}
\usepackage{epsfig}

\newcommand{\be}{\begin{equation}}
\newcommand{\ee}{\end{equation}}
\newcommand{\beq}{\begin{eqnarray}}
\newcommand{\eeq}{\end{eqnarray}}

\newcommand{\aeta}[3]{  #1, {\it Astron. Astrophys.}, {\bf  #2}, #3.}

\newcommand{\aspj}[3]{  #1, {\it Astrophys. J.}, {\bf  #2}, #3.}

\newcommand{\sph}[3]{   #1, {\it Solar Phys.}, {\bf  #2}, #3.}

\begin{document}
\begin{article}
\begin{opening}

\title{COMPARISON OF LARGE-SCALE FLOWS ON THE SUN MEASURED BY TIME-DISTANCE HELIOSEISMOLOGY
AND LOCAL CORRELATION TRACKING TECHNIQUE}

\author{Michal \surname{\v SVANDA}$^{1,2}$,
        Junwei \surname{ZHAO}$^{3}$,
        Alexander G. \surname{KOSOVICHEV}$^{3}$}

\runningauthor{\v SVANDA {\it et al.}} \runningtitle{Large-scale flows
from time-distance helioseismology and LCT}

\institute{$^{1}$ Astronomical Institute of Academy of Sciences,
Ond\v rejov Observatory, Ond\v rejov, CZ-251~65, Czech Republic
                  \email{email: svanda@asu.cas.cz}\\
           $^{2}$ Astronomical Institute, Charles University, V Hole\v sovi\v ck\'ach 2, Prague, CZ-182~00, Czech Republic\\
           $^{3}$ W.~W.~Hansen Experimental Physics Laboratory, Stanford University, Stanford, CA94305-4085, USA}

\date{Received 10 October 2006, accepted 23 January 2007}

\begin{abstract}

\noindent We present a direct comparison between two different
techniques time-distance helioseismology and a local correlation
tracking method for measuring mass flows in the solar photosphere
and in a near-surface layer: We applied both methods to the same
dataset (MDI high-cadence Dopplergrams covering almost the entire
Carrington rotation 1974) and compared the results. We found
that after necessary corrections, the vector flow fields obtained by
these techniques are very similar. The median difference between
directions of corresponding vectors is 24\,$^\circ$, and the
correlation coefficients of the results for mean zonal and meridional
flows are 0.98 and  0.88 respectively. The largest discrepancies are
found in areas of small velocities where the inaccuracies of 
the computed vectors play a significant role. The good agreement of
these two methods increases confidence in the reliability of
large-scale synoptic maps obtained by them.

\end{abstract}
\keywords{Sun: surface flows, time-distance helioseismology, local
correlation tracking}

\end{opening}

\section{Motivation}

There are several methods for measuring flows in surface layers of
the Sun. Local helioseismology uses information about solar
oscillations\,--\,frequency shifts and travel time variations\,--\,to infer
the structure of solar interior and to determine flow patterns just
below the surface (\emph{e.g.} \citeauthor{zhao04a}, \citeyear{zhao04a}).
The local correlation tracking method measures apparent motion of
specific structures to determine the flow field (\emph{e.g.} motions of
granules -- \citeauthor{sobotka99}, \citeyear{sobotka99}, or
magnetic elements -- \citeauthor{meunier2005}, \citeyear{meunier2005}). A similar method is feature
tracking, which evaluates motions of well defined isolated features.
Direct Doppler measurements provide in general just the
line-of-sight component of the velocity vector, but when applied to
large datasets, these can provide statistically robust information
about properties of the surface flows (\emph{e.g.} the discovery of
supergranulation by \citeauthor{hart56}, \citeyear{hart56} or
\citeauthor{leighton62}, \citeyear{leighton62}).

However, the results obtained by different methods may have
discrepancies. These can be caused by the nature of the methods,
\emph{e.g.} due to different types of averaging, and also because of the
use of different datasets from different instruments. In addition,
various disturbing effects may be important. Therefore, we decided
to compare the results obtained by two different methods,
time-distance helioseismology and local correlation tracking (LCT),
using the same set of data: high-cadence Dopplergrams, covering
almost one Carrington rotation, obtained from Michelson Doppler
Imager (MDI, \citeauthor{scherrer95}, \citeyear{scherrer95}).

Solar acoustic waves ($p$~modes) are excited in the upper convection
zone and travel between various points on the surface through the
interior. The travel time of acoustic waves is affected by
variations of the speed of sound along the propagation paths and
also by mass flows. Time-distance helioseismology measurements
\cite{duvall93} and inversions (\citeauthor{kosovichev96},
\citeyear{kosovichev96}, \citeauthor{zhao01}, \citeyear{zhao01})
provide a tool to study three-dimensional flow fields in the upper
part of the solar convection zone with relatively high spatial
resolution.

The local correlation tracking (LCT) method was originally designed
for removing  seeing-induced distortions in sequences of solar
images \cite{november86}, and later used for mapping motions of
granules in series of white-light images of the photosphere
\cite{november88}. This method works on the principle of best match:
local displacements in the image frames are determined for each
position by cross-correlating pairs of sub-frames within a
pre-defined spatial window (correlation window), and then the
corresponding velocities are calculated from these displacements.

\label{svanda:lct_under}
In some studies it has been shown that LCT technique underestimates the real velocities
due to the smoothing of processed data by the correlation window. For
example, in the study by \citeauthor{svanda06} (\citeyear{svanda06}) based on synthetic Dopplergrams, it was found
that the LCT code used also in this study underrepresents the magnitudes of velocities
by a factor of 1.13. In \citeauthor{georgobiani06} (\citeyear{georgobiani06}), the correction factor with the same
meaning was found to be 1.5. However, Georgobiani's study was
done using a another LCT code applied to simulated data and the
results were compared with time-distance method applied to $f$~modes.
This means that the correction factor is different for different
parameters used in the LCT method and, therefore, it should be determined
(calibrated) empirically for each particular study. Application of LCT to MDI Dopplergrams 
by \citeauthor{derosa04} (\citeyear{derosa04})  showed the underestimation by 
more than 30\%. Other studies also find the 
underrepresenting of magnitudes by LCT (\citeauthor{sobotka99}, \citeyear{sobotka99} -- 20~\%, 
\citeauthor{roudier99}, \citeyear{roudier99} -- 25~\%).

Both methods provide surface or near-surface velocity vector fields.
However, the results of these methods can be interpreted
differently. While local helioseismology measures intrinsic plasma
motions (through advection of acoustic waves), LCT measures apparent
motions of structures (granules or magnetic elements). It is
known that some structures do not necessarily follow the flows of
the plasma on the surface. Fox example, supergranulation appears to rotate
faster than the plasma (\citeauthor{beck00}, \citeyear{beck00}), which may be
caused by travelling waves (\citeauthor{gizon03}, \citeyear{gizon03}) or may 
be explained also as a projection effect (\citeauthor{hathaway06}, \citeyear{hathaway06}). 
Some older studies (\emph{e.g.} \citeauthor{rhodes91}, \citeyear{rhodes91}) also suggest 
that the difference in flow properties measured on the basis of structures' motions and 
plasma motions is caused by a deeper anchor depth of these structures. An evolution of the pattern 
may also play a significant role (\emph{e.g.} due to emergence of
magnetic elements). Another possiblity is
that surface structures are not coherent features, but patterns 
traveling with a different group velocity than the surface plasma 
velocity, such as occurs for the features present in simulations of 
travelling-wave convection (\emph{e.g.}, \citeauthor{hurlburt96}, \citeyear{hurlburt96}).

Some attempts to compare the results of local helioseismology and the
LCT method for large scales, with characteristic size 100~Mm and
more, have been carried out by \citeauthor{ambroz05}
(\citeyear{ambroz05}), but his results were inconclusive. The
correlation coefficient describing the match of the velocity maps
obtained by local helioseismology and the LCT method was close to
zero. Nevertheless, there were compact and continuous regions of
characteristic size from 30 to 60 heliographic degrees with a good
agreement between the two methods, so that one could not conclude
that the results were completely different. In his study, many
factors could be significant: the techniques were applied to
different types of datasets (LCT was applied to low resolution
magnetograms acquired at the Wilcox Solar Observatory and the
time-distance method used MDI Dopplergrams). Both techniques had
very different spatial resolution, and also the accuracy of the
measurements was not well known.

We decided to avoid these problems and analyze the same data set
from the MDI instrument on SOHO. MDI provides approximately two
months of continuous high-cadence (one-minute cadence) full-disc
Dopplergrams each year. This \emph{Dynamics Program} provides data
suitable for helioseismic studies, and also for the local correlation
tracking of supergranules. Thus, this is a perfect opportunity to
compare the performance and results of two different techniques using
the same set of data, and to avoid effects of observations with
different instruments or in different conditions.

\section{Data preparation}

The selected dataset consists of 27 data-cubes from March 12,
2001, 0:00~UT to April 6, 2001, 0:31~UT, where each third day was
used, and in these days three 8.5-hour long data-cubes were
processed (so that every third day in the described interval was
fully covered by measurements). Each data-cube is composed of 512
Dopplergrams with spatial resolution of 1.98$^{\prime\prime}\,\rm px^{-1}$ 
at a one-minute cadence. All of the frames of each data-cube were tracked with rigid
rate of 2.871~$\mu$rad\,s$^{-1}$, remapped with Postel's
projection with a resolution of 0.12\,$^\circ$\,px$^{-1}$ 
(1\,500~km\,px$^{-1}$ at the center of the disc), 
and only a central meridian region was selected for further processing
(with size of 256$\times$924~px covering $30\,^\circ$ in
longitude and running from $-54\,^\circ$ to $+54\,^\circ$ in
latitude), so that effects of distortions due to the projection do
not play a significant role.

Tracked data-cubes were used to perform the time-distance analysis.
From all the frames in each data-cube, the mean Dopplergram (like
Figure~\ref{examples} left) was subtracted to suppress the influence
of velocity structures like supergranulation and to highlight the
signals of $p$~modes. $P$~modes have their origin in the 
solar convection zone and travel through the solar interior
to the surface. The travel-time of the wave depends 
on the speed of sound and on the velocity of the mass flow
in the layers of the solar interior, through which the bulk is travelling.
In the time-distance technique, the travel times of waves from 
the point in the photosphere (central point) to a surrounding annuli around 
this point are measured. The radius of each annulus selects wavemodes that propagate down to a
specific depth, before being refracted upward toward the photosphere. Travel times are measured by the cross-correlation between Doppler velocities in the central point and velocities in the selected annuli around this point. 

The mass flow velocities in the interior are calculated from the differences
of travel times from the central point to the surrounding annuli and the travel 
times from the surrounding annuli to the central point when the state properties 
in the affected layers of the solar interior are known. In this study, the theoretical
ray approximation is derived from the solar model S (\citeauthor{christensen96}, \citeyear{christensen96}). 
Dividing the annuli into sectors, the underlying flow field of selected orientations can be infered.
For details see \citeauthor{kosovichev96} (\citeyear{kosovichev96}), \citeauthor{zhao01} 
(\citeyear{zhao01}), or \citeauthor{zhao04a} (\citeyear{zhao04a}).

During the process, the surface gravity wave ($f$~mode) is filtered out from the
$k$--$\omega$ diagram before computing the travel times, because it has different dispersion characteristics than the $p$~modes
used in this study. 
$F$~modes, if not filtered out, will disturb $p$~modes measurements, and it is also not
straightforward to perfrom inversions if not separating two different modes. The $p$~mode 
inversions are less sensitive to the surface flows than $f$~mode data, but still recover the 
large-scale flows well.

The  time-distance inversion results were smoothed by a Gaussian
with FWHM of 30~px to match the resolution to the LCT method, and
only the horizontal components ($v_x$, $v_y$) of the full velocity
vector were used.

While for the time-distance method the $p$~modes of solar
oscillations play a crucial role, they significantly influence the
performance of the LCT method in a negative way. The oscillations
are clearly visible in the Dopplergrams, causing random
errors in the calculation of displacements. Therefore, before
applying the LCT method, the oscillation signals must be suppressed.
For our high-cadence data it is possible to do this using  temporal
averaging. According to \citeauthor{hathaway88}
(\citeyear{hathaway88}) or more recently \citeauthor{hathaway00}
(\citeyear{hathaway00}) it is better to use a Gaussian type of
temporal averaging than the boxcar one. We average the Dopplergrams
over a 31-minute period with weights given by the formula
\begin{equation}
w(\Delta t)=  \exp\left[{\frac{(\Delta
t)^2}{2a^2}}\right]-\exp\left[{\frac{b^2}{2a^2}}\right] \left(1+\frac{b^2-(\Delta
t)^2}{2a^2}\right),
\end{equation}
where $\Delta t$ is the time between a given frame and the central one
(in minutes), $b=16$~minutes and $a=8$~minutes. We verified that this filter
suppresses the solar oscillations in the 2\,--\,4 mHz frequency band by
a factor of more than five hundred.

The other issue significantly influencing the performance of the LCT
 method is the change of  contrast and background intensity caused by
solar rotation. Due to tracking the Doppler images with rotation,
the magnitude of the line-of-sight component of the solar rotational
velocity changes from frame to frame and affects the LCT results.
The method interprets these changes as a motion towards the East,
mainly in the central part of the solar disc, where the contrast in
the structures of Dopplergrams is very low (see Figure~\ref{examples}
left). We suppress the influence of the moving background by
subtraction of a third-order polynomial surface fit. We have
tested that this provides almost the same results as the other
possible procedures: local removal of the mean values and unsharp
masking. Subtraction of the polynomial fit is not so sensitive
to anomalies in the Dopplergrams, caused by regions with strong
magnetic field.

The LCT method used in this study is described by the following
parameters: the time-lag between correlated frames is 120 frames (two
hours), the correlation window has a Gaussian shape with FWHM of 30
px, the correlation is measured by the sum of absolute differences
of subframes (it is faster than calculation of the correlation
coefficient and provides the same results), the extremum position
is calculated using the nine-point method \cite{darvann91}. For each
data-cube, the results of all the correlated pairs are averaged, so
that the method provides an averaged flow field over 8.5~hours in the
same sense as the time-distance analysis.

\section{Results}

\begin{figure}[!t]
\begin{center}
\psfig{file=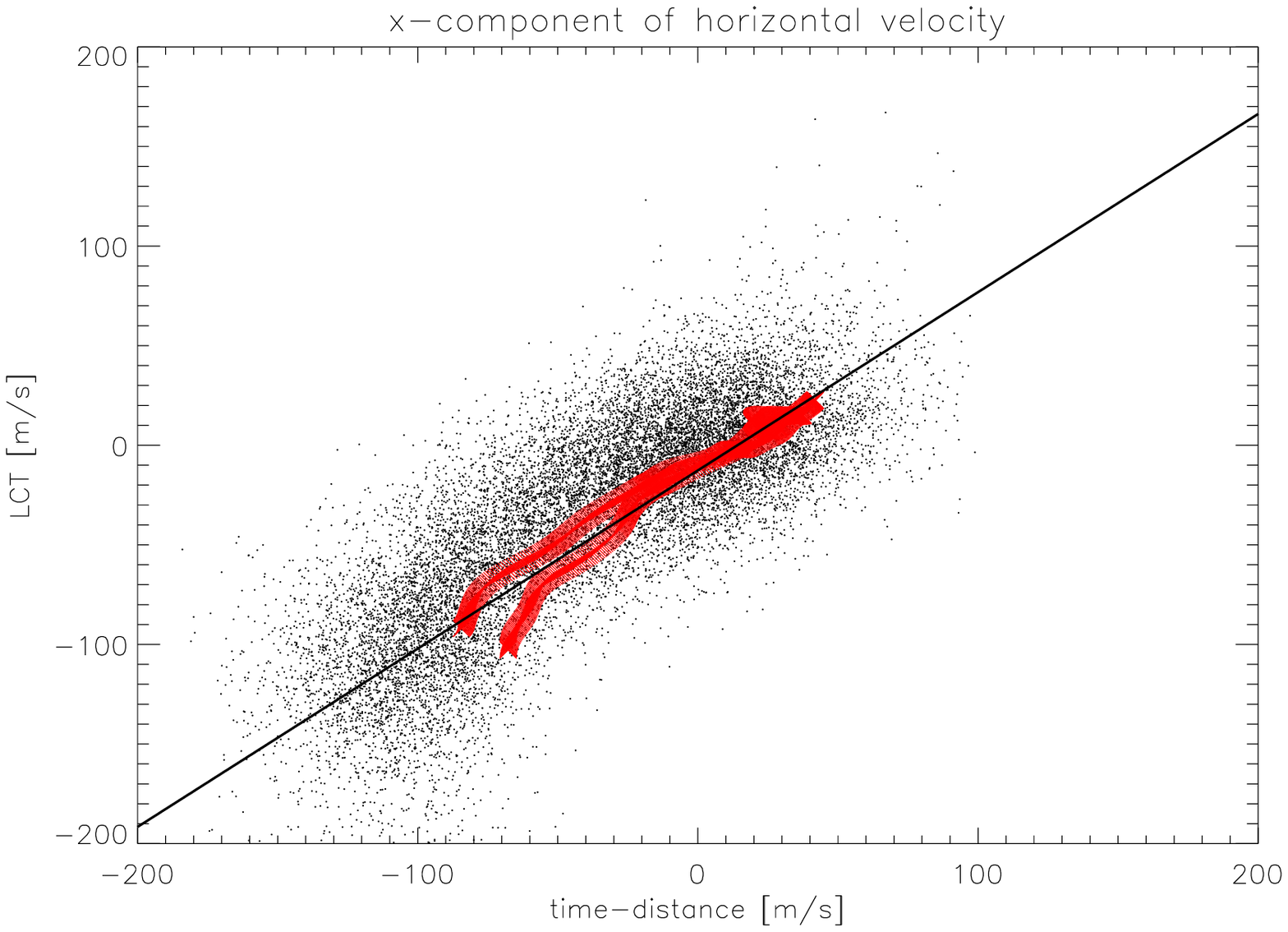,width=0.49\textwidth,clip=}
\psfig{file=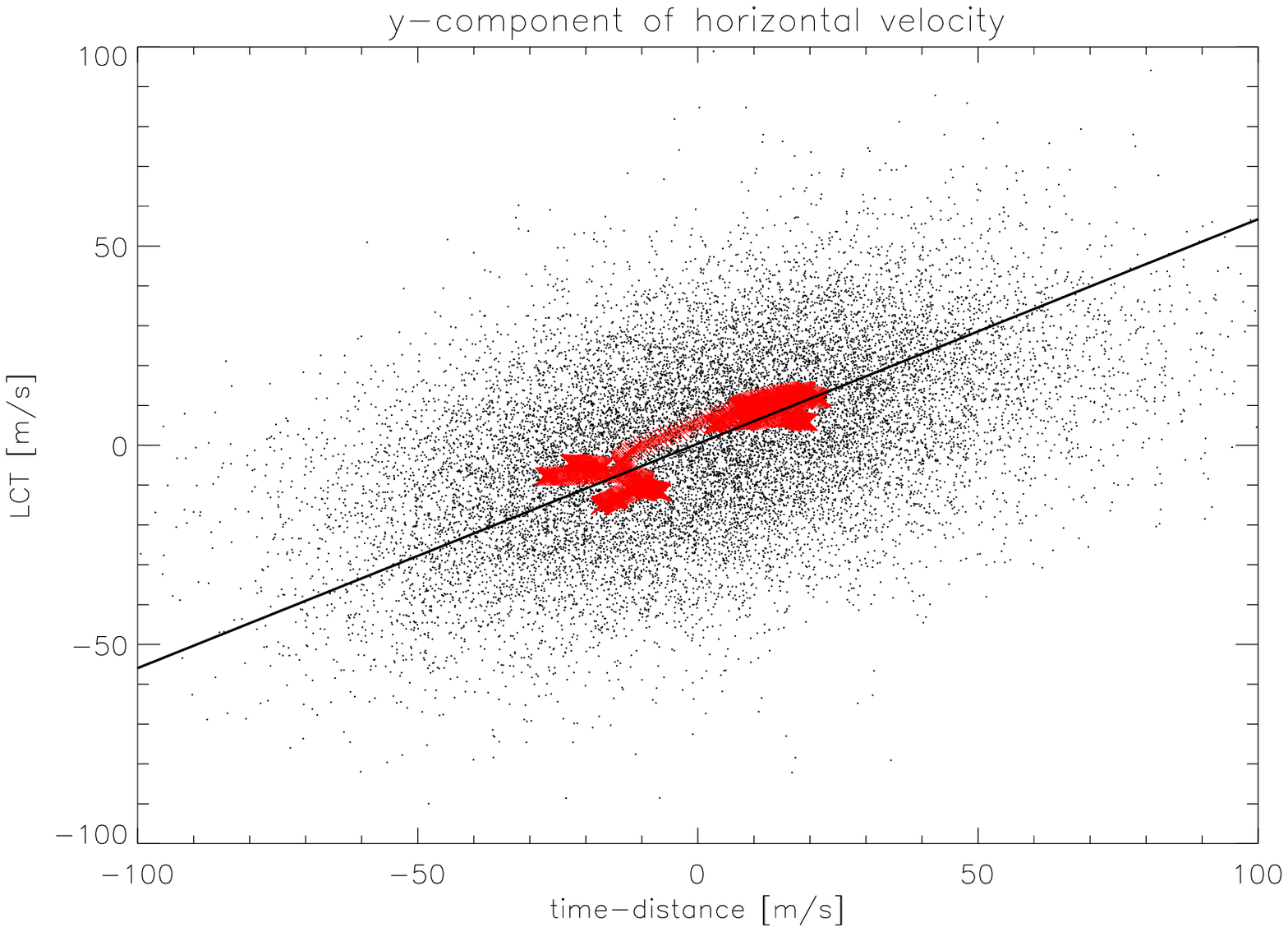,width=0.49\textwidth,clip=}
\caption{\emph{Left} -- $v_{x,{~\rm LCT}}$ {\it versus} $v_{x,{~\rm t-d}}$
plot. Red crosses denote mean zonal velocities (differential
rotation) which have been used for  fitting of the calibration
Equation (\ref{vx_fit}). \emph{Right} -- the same for $v_y$
component, the regression fit is described by Equation (\ref{vy_fit}).}
\label{vx,vy}
\end{center}
\end{figure}

\subsection{Statistical processing}
\label{statistics}

The results containing 27 horizontal flow fields were statistically
processed to obtain the cross-calibration curves for these methods.
It is generally known (see discussion in Section 1, page~\pageref{svanda:lct_under})
that the LCT method slightly underestimates the
velocities; thus, the results should be corrected by a certain
factor. From the comparison of the $x$-component of velocity (cf.
Figure~\ref{vx,vy} left) we obtained parameters of a linear fit given
by (numbers in parentheses denote a $\sigma$-error of the regression
coefficient)
\begin{equation}
v_{x,{~\rm LCT}}=0.895(0.008) v_{x,{~\rm t-d}}-12.6(0.3)~{\rm m\,s^{-1}} .
\label{vx_fit}
\end{equation}
The correlation coefficient between $v_{x,{~\rm LCT}}$ and
$v_{x,{~\rm t-d}}$ is $\rho=0.80$. We assume that the time-distance
measurements for $v_{x,{~\rm t-d}}$ are correct and the magnitude
of the LCT measurements, $v_{x,{~\rm LCT}}$, must be corrected
according to the slope of Equation~(\ref{vx_fit}). This correction factor
has a value of 1.12, which is in perfect agreement with the
correction factor of 1.13 found in the tests of the same LCT code
using synthetic Dopplergrams with the same resolution and similar
LCT parameters \cite{svanda06}. We assume that both velocity
components obtained with the LCT method should be corrected by this
factor. 

The regression line of $v_y$ component (Figure~\ref{vx,vy} right)
is
\begin{equation}
v_{y,{~\rm LCT}}=0.56(0.01) v_{y,{~\rm t-d}}+0.4(0.2)~{\rm m\,s^{-1}} .
\label{vy_fit}
\end{equation}
After the slope correction using the $v_x$ fits, the regression
curve is slightly different:
\begin{equation}
v_{y,{~\rm LCT}}=0.63(0.01) v_{y,{~\rm t-d}}+0.4(0.2)~{\rm m\,s^{-1}},
\end{equation}
with the correlation coefficient between $v_{y,{~\rm LCT}}$ and
$v_{y,{~\rm t-d}}$ close to 0.47. The slope of the linear fit
differs significantly from the expected value 1.0. 

We have tested that this asymmetry is not related to the LCT
technique. The tests did not show any preference in direction of flows
measured by LCT or any dependence of the results on the size
of the field of view (which is not a square). Also in the
recent study by \citeauthor{svanda06} (\citeyear{svanda06}), based 
on synthetic data, the asymmetry between the zonal and the meridional
components was not encountered. 

It is possible that a drift of the supergranular
pattern towards the equator (such as found by \citeauthor{gizon03}, \citeyear{{gizon03}}) might cause an asymmetry
between measurements of the north-south and east-west
velocities.  However, this process does not explain why the meridional
velocities from both techniques seem to be proportional to each
other. A systematic drift would rather be depicted as a
constant, or a shift depending on the latitude. However, the
meridional components of velocities are generally rather small, so
the errors of the measurements can play a significant role and
the proportional behavior can be only apparent. Such an effect is not seen
in the experiments with test data, so we do not favor this explanation.

A second
explanation is based on unspecified asymmetries influencing travel-time
measurements, for instance, due to different sensitivity of the MDI
instrument to $p$~modes propagating in the East\,--\,West and North\,--\,South
directions. \citeauthor{georgobiani06} (\citeyear{georgobiani06}) did not find such an asymmetry 
using $f$-mode time-distance and LCT applied to
realistic numerical simulation. The asymmetry in the East\,--\,West and
North\,--\,South directions observed by $p$~mode time-distance helioseismology
was on the contrary noticed in a recent study based on numerically simulated
data (\citeauthor{zhao06}, \citeyear{zhao06}). This evidence suggests the asymmetry arises only in $p$~mode inversions such
as those studied here, and should be
investigated further in more detail.

In this study we
decided to correct the $y$-component of the time-distance results.
\begin{figure}[!t]
\begin{center}
\psfig{file=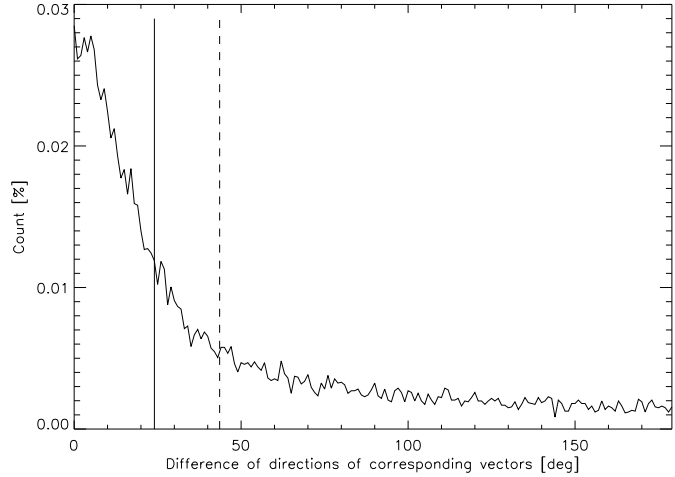,width=0.75\textwidth,clip=} \caption{Histogram
of the angular differences ($\Delta \varphi$) between directions of
the velocity vectors obtained by the time-distance and LCT
techniques. Dashed vertical line denotes the mean value and solid
vertical line represents the median of $\Delta \varphi$.} \label{histy}
\end{center}
\end{figure}
The final calibration formulae providing the best statistical
agreement between the velocities calculated using both methods are:

\begin{eqnarray}
v_{x, \rm LCT, corr} & = & 1.12\,v_{x, \rm LCT, calc} \label{calbeg}\\
v_{y, \rm LCT, corr} & = & 1.12\,v_{y, \rm LCT, calc} \\
v_{x, \rm t-d, corr} & = & v_{x, \rm t-d, calc} \\
v_{y, \rm t-d, corr} & = & 0.63\,v_{y, \rm t-d, calc}\label{calend},
\end{eqnarray}
where the index \emph{corr} denotes the corrected value, and the index
\emph{calc} denotes the original calculated value.

After the corrections, as presented in the histogram in
Figure~\ref{histy}, the differences between the directions of the
velocity vectors ($\Delta \varphi$) calculated by these techniques are quite
reasonable. The mean value of the distribution is 43.56\,$^\circ$,
however, the mean value is not a good indicator in this case because
the distribution function is not normal. The median value is
24.02\,$^\circ$ and 66.6~\% of points have the difference in the
corresponding vector directions less than 45\,$^\circ$. 

Instead of computation of the correlation coefficient of the arguments 
of both vector fields, we decided to compute a magnitude-weighted cosine 
of $\Delta \varphi$, because it provides more robust results. This quantity is given by 
\begin{equation}
\rho_{\rm W} = \frac{\sum |{\mathbf v}_{t-d}| \frac{|{\mathbf v}_{t-d} \cdot {\mathbf v}_{LCT}|}
{|{\mathbf v}_{t-d}||{\mathbf v}_{LCT}|}}{\sum |{\mathbf v}_{t-d}|}, 
\end{equation}
where ${\mathbf v}_{t-d}$ is the time-distance vector field, ${\mathbf v}_{LCT}$ is the LCT vector field
and the summation is performed over all vectors in the field. The closer this quantity is to one, the 
better is the agreement between both vector fields. Larger vectors are weighted more than smaller ones. 
We have found that in our case $\rho_{\rm W}=0.86$.

\subsection{Mean velocities}
\begin{figure}[!t]
\begin{center}
\psfig{file=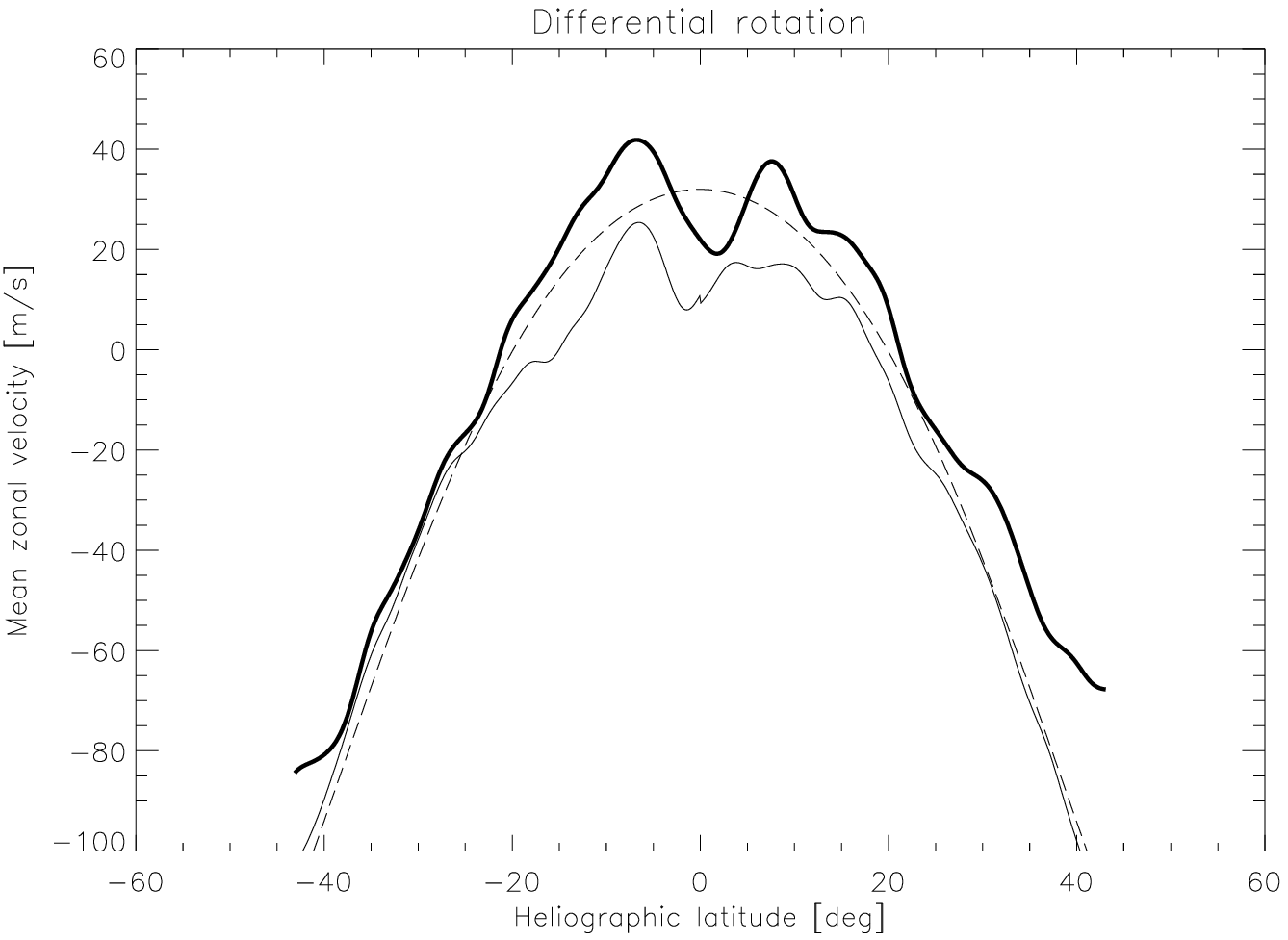,width=0.49\textwidth,clip=}
\psfig{file=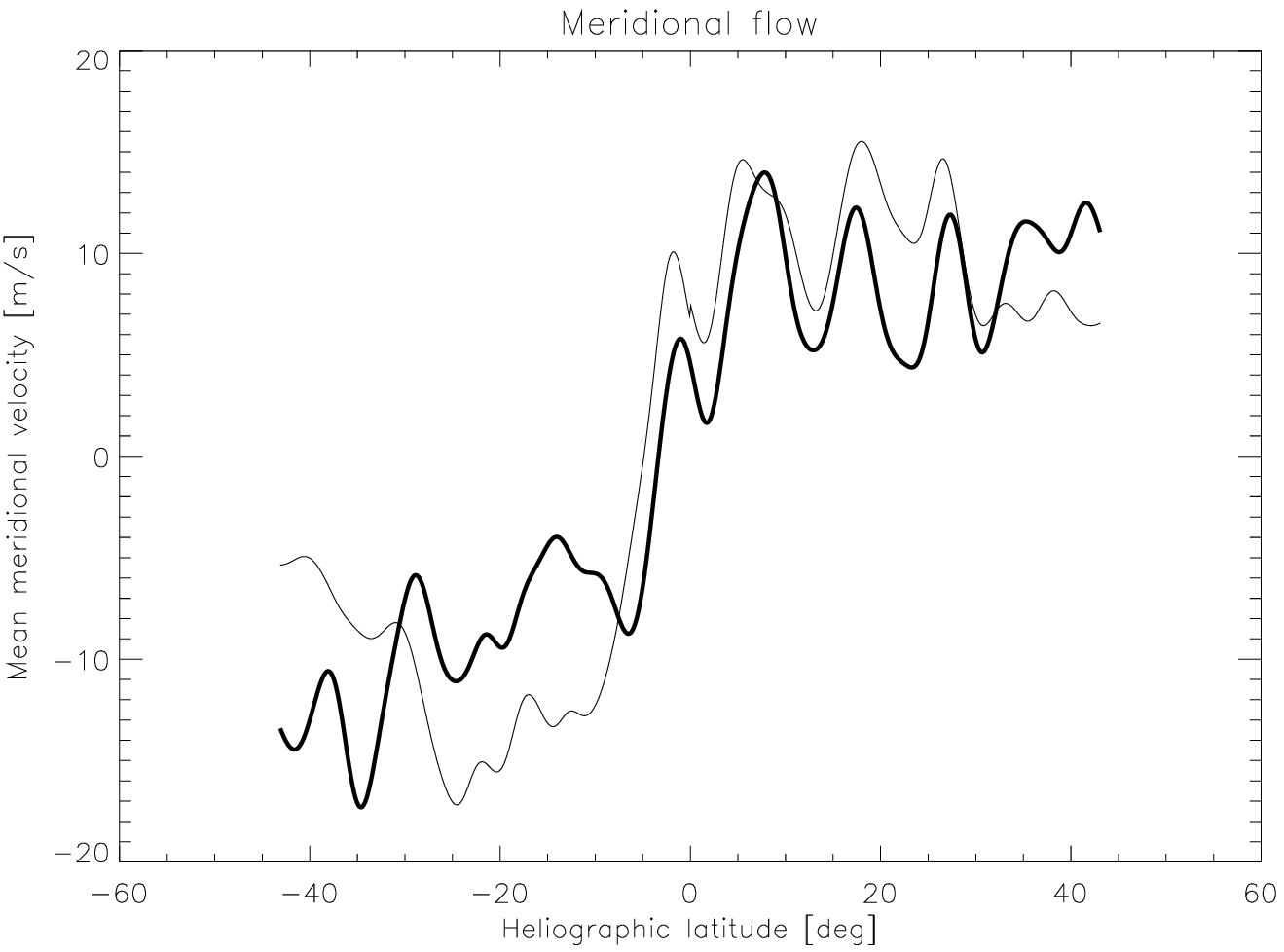,width=0.49\textwidth,clip=} \caption{{\it
Left:} Mean zonal velocity (thick curve) as a function of
latitude (differential rotation) obtained from the time-distance
data and mean zonal velocity (thin curve) obtained from the LCT
data. Dashed line represents a standard rotation profile
(``Snodgrass rate''). The mean zonal velocities are plotted in the
coordinate system rotating rigidly with 2.871~$\mu$rad\,s$^{-1}$.
{\it Right:} Velocities of the mean meridional flow as a function of
latitude obtained by the time-distance technique (thick curve) and
by the LCT method (thin curve). The correlation coefficient for the
mean zonal flow is $\rho=0.98$ and for the mean meridional flow
$\rho=0.88$. } \label{mean}
\end{center}
\end{figure}
\begin{figure}[!t]
\begin{center}
\psfig{file=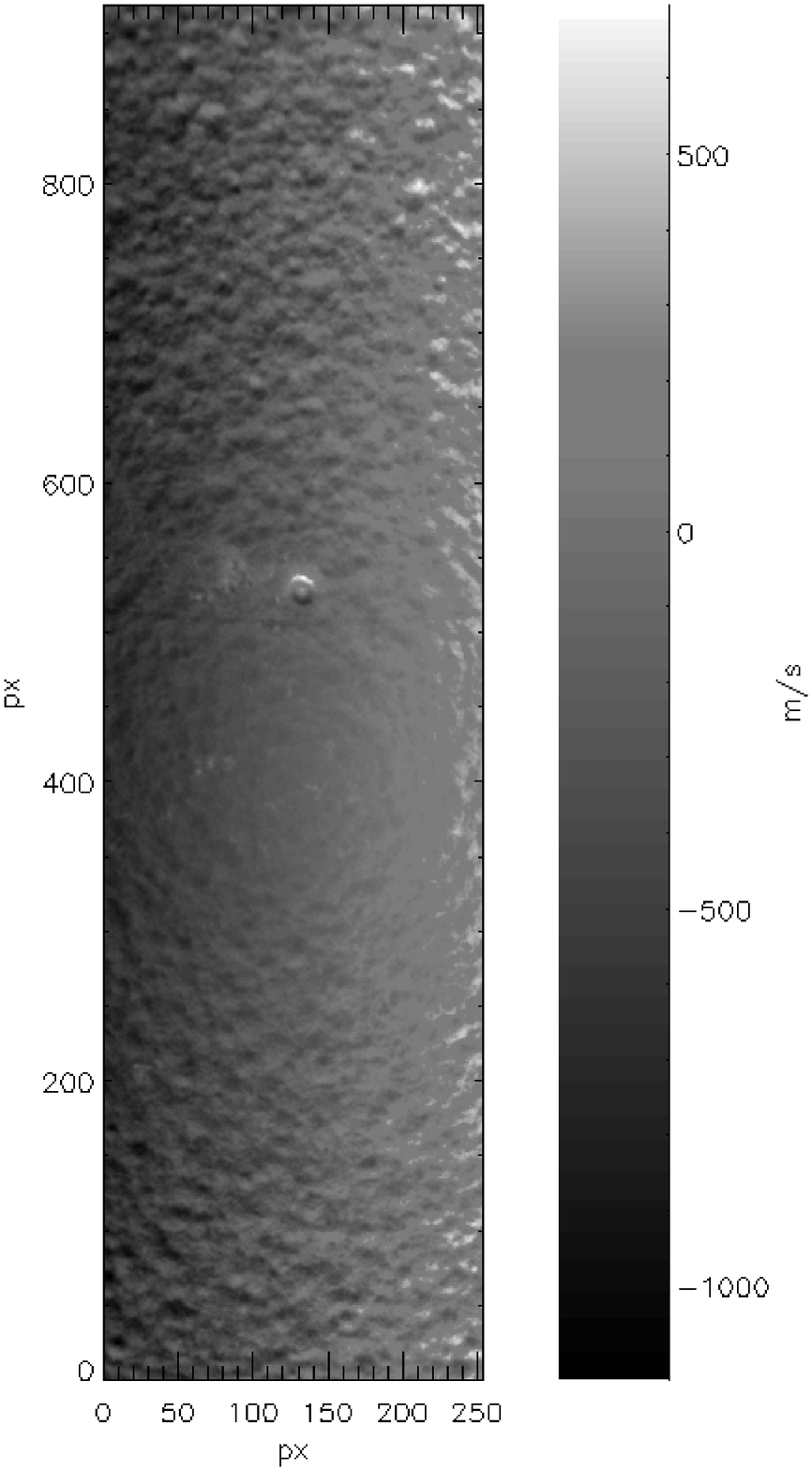,height=10cm,clip=} \rule{0.5cm}{0pt}
\psfig{file=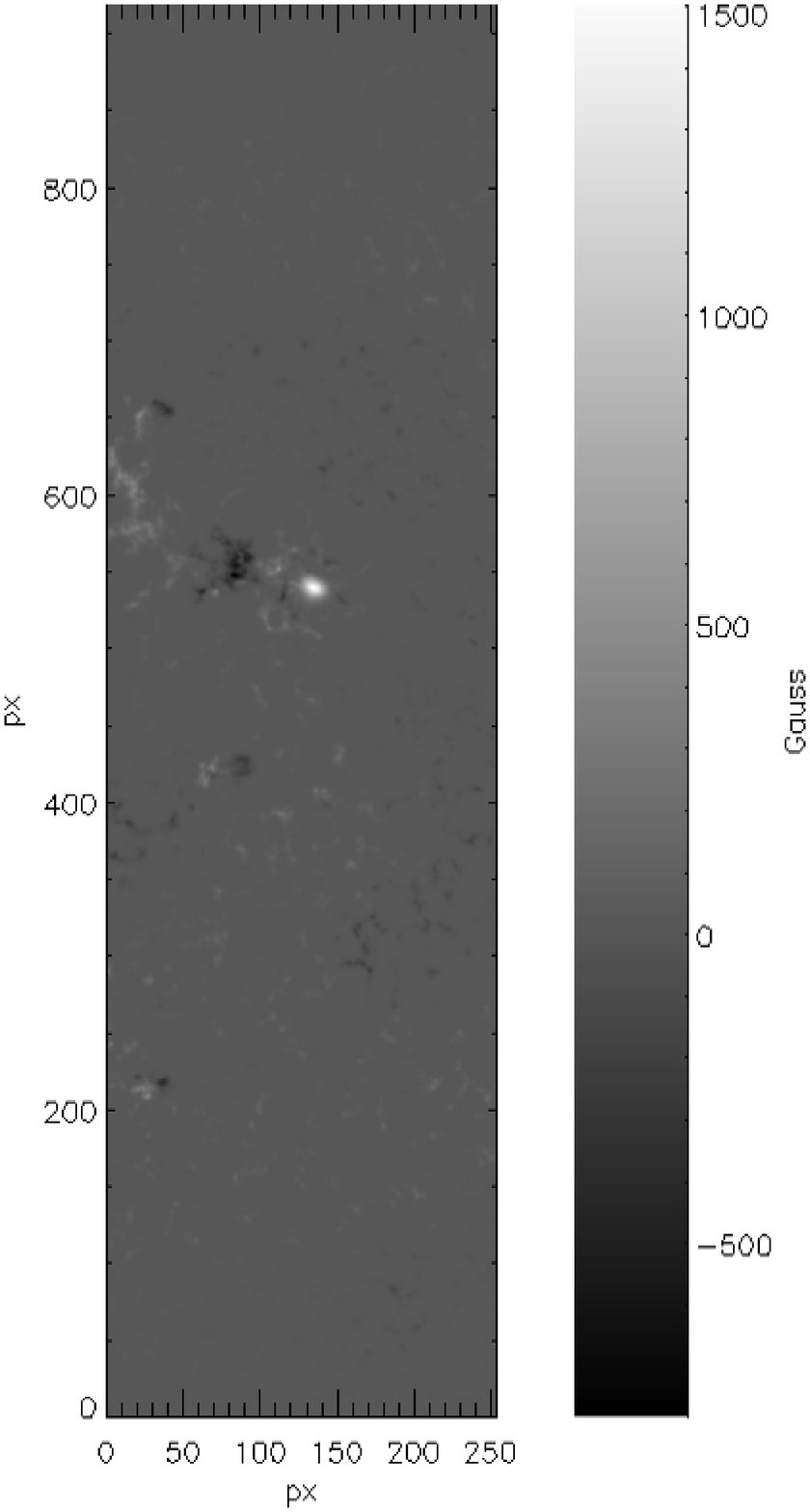,height=10cm,clip=} \caption{Part of the solar
disc chosen for the detailed comparison of the results of
time-distance and LCT methods. The 8.5-hour measurements are centered at
March 24 2001, 4.16~UT.  One pixel represents 0.12\,$^\circ$ in
heliographic coordinates. \emph{Left} -- Averaged MDI Dopplergram.
Note the low contrast of the supergranular cells in the center of
the image. This is the ``blind spot'' caused by prevailing
horizontal motions in the photosphere.
\emph{Right} -- Averaged MDI magnetogram.} \label{examples}
\end{center}
\end{figure}

In addition to the detailed comparison of the vector fields, we
compare the mean flows, the differential rotation, and the meridional
circulation. These flows can be quite simply calculated from the
results of both techniques. In both cases, they provide the mean
zonal and mean meridional flows for Carrington rotation
1974. The results are displayed in Figure~\ref{mean}, where the
differential rotation curves are compared with a standard profile
from \citeauthor{snodgrass90} (\citeyear{snodgrass90}) in the left panel. It can be clearly seen
that the latitudinal profiles for both techniques are very similar
and also that the mean velocities do not differ much in
magnitude. The correlation coefficients are $\rho=0.98$ for the
zonal flow and $\rho=0.88$ for the meridional flow. In the
differential rotation curves, the LCT results give a little slower
rotation, which is also seen from Equation~(\ref{vx_fit}). The mean difference
of average zonal velocities obtained by both techniques is 14.1~m\,s$^{-1}$.

We conclude that for the mean flows the results obtained by the techniques agree very well.

\subsection{Detailed comparison}

\begin{figure}[!t]
\begin{center}
\psfig{file=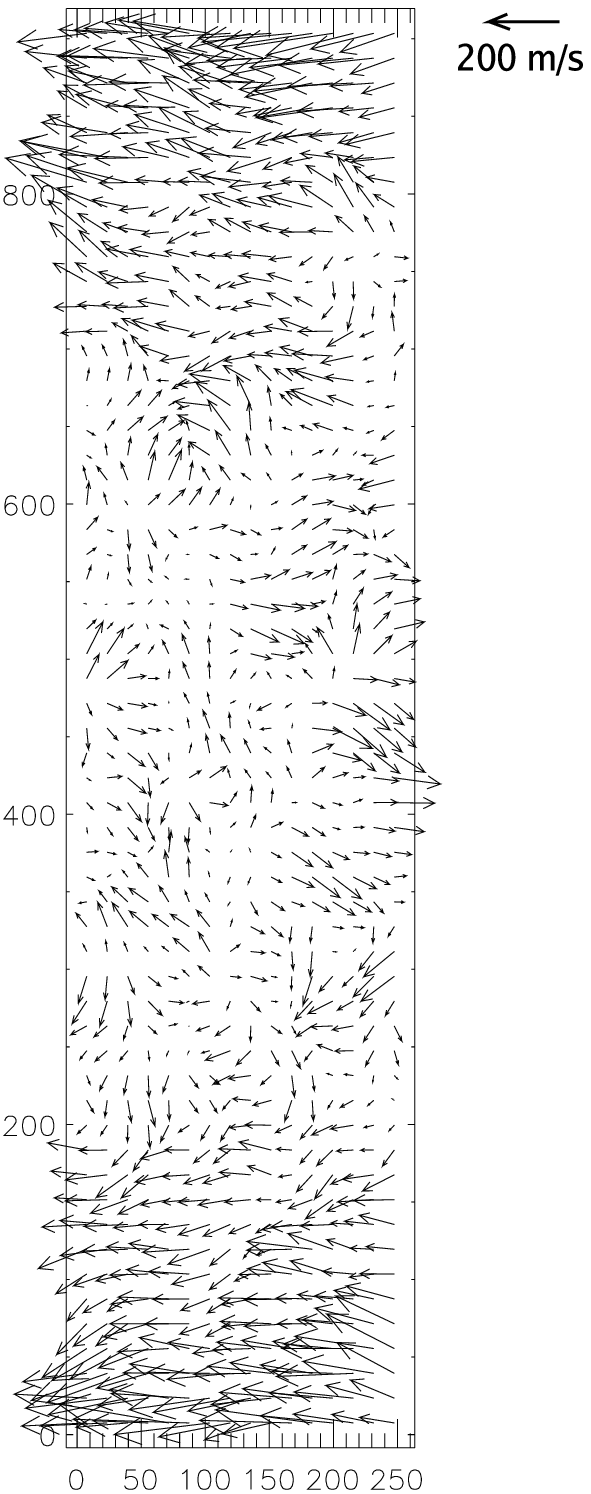,height=12cm,clip=}
\psfig{file=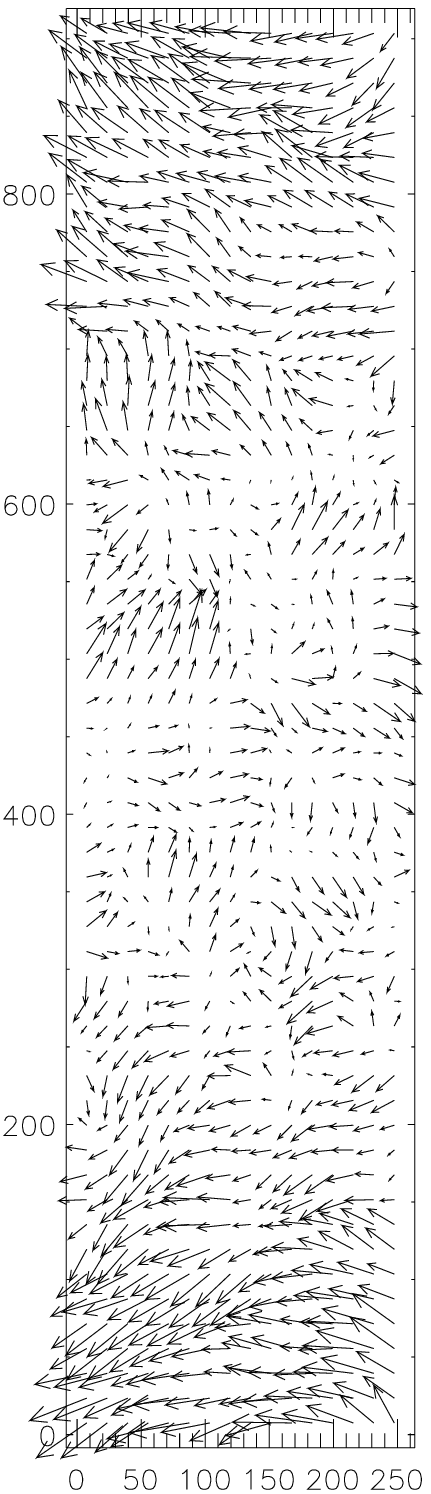,height=12cm,clip=} \caption{\emph{Left} --
Velocity field obtained by the LCT method. \emph{Right} -- Velocity
field obtained by the time-distance technique. Both plots are
centered at heliographic coordinates $b_0=0.0\,^\circ$,
$l_0=214.3\,^\circ$. Units on both axes are pixels in the data frame
with resolution of 0.12\,$^\circ$\,px$^{-1}$ in the Postel
projection.} \label{arrows}
\end{center}
\end{figure}
\begin{figure}[!t]
\begin{center}
\psfig{file=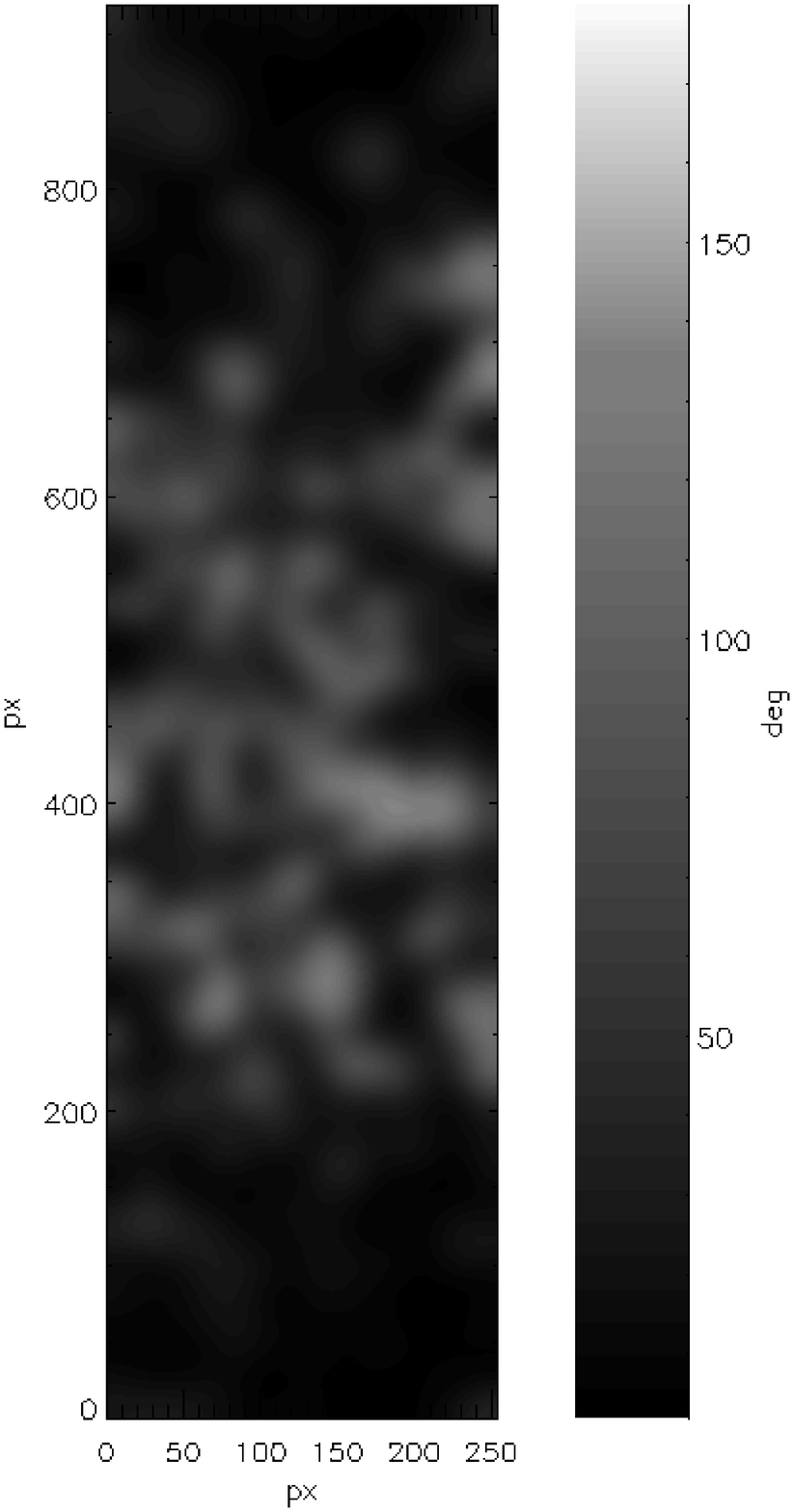,height=10cm,clip=}
\psfig{file=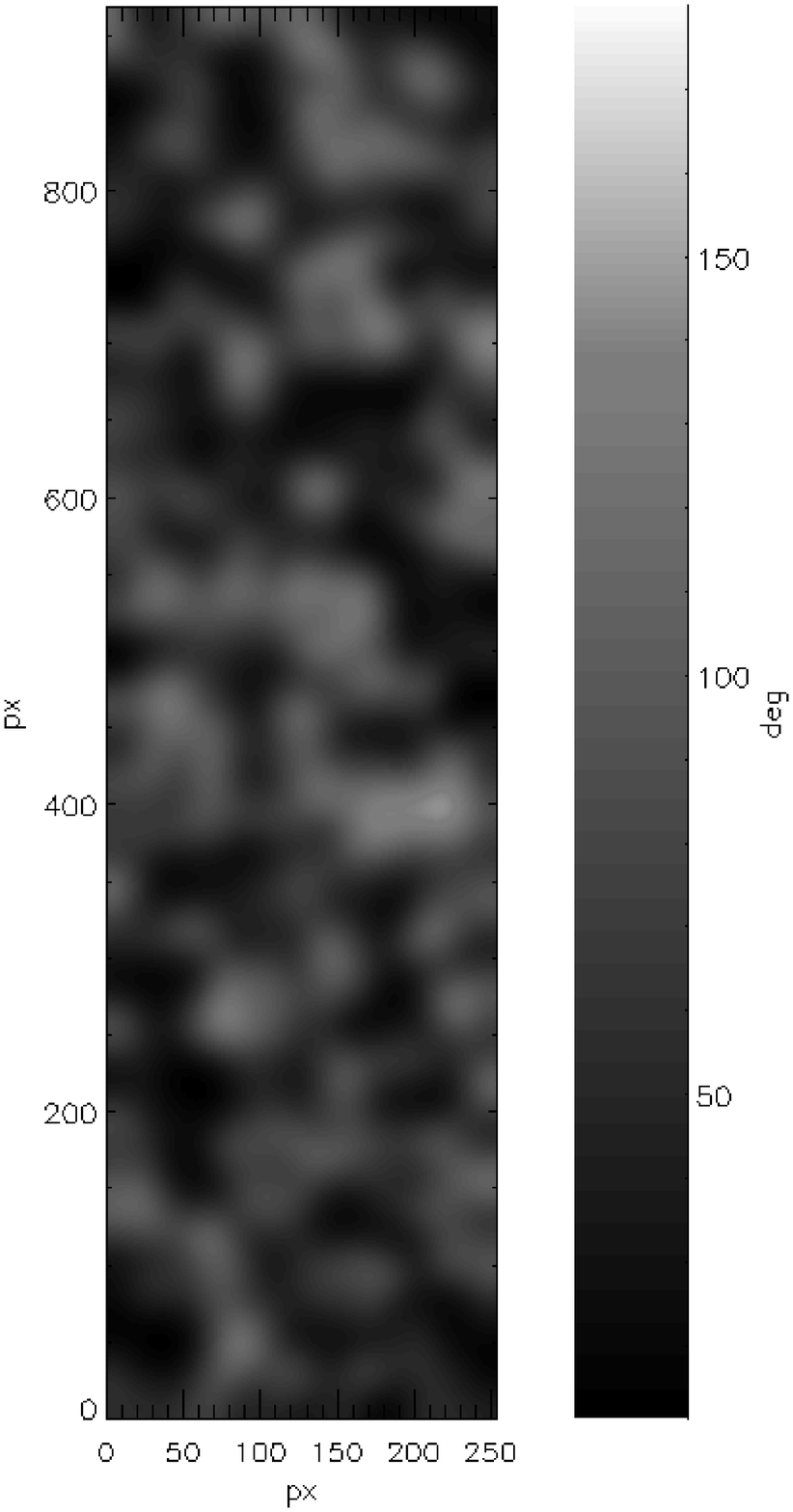,height=10cm,clip=} \caption{Difference of
phases of corresponding vectors ($\Delta\varphi$). \emph{Left} -- for the full vector data, \emph{right} -- for the data with mean zonal flow substracted. In the second case the magnitudes of vectors everywhere in the studied area are comparable and the inaccuracies of individual measurements equally important. With the mean zonal flow removed the distribution of $\Delta\varphi$ become more uniform in the studied area.} \label{difphases}
\end{center}
\end{figure}

For a detailed comparison of the flow fields, we selected one data
cube, representing 8.5-hour measurements centered at 4:16~UT March
24, 2001, $l_0$=214.3\,$^\circ$ (see the averaged MDI Dopplergram
and magnetogram in Figure~\ref{examples}). In this map, the
correlation coefficient for the $x$-component of the velocity is
$\rho=0.82$, for the $y$-component $\rho=0.58$, and for the
vector magnitude: $\rho=0.73$. The vector plots of the flow fields
obtained by both techniques, shown in Figure~\ref{arrows}, in general seem to be
quite similar to each other. However, many differences can be seen. The
regions  where the differences are most significant correspond to
relatively small (under 50~m\,s$^{-1}$) velocities. This is clear
from the map of the differences between the vector directions,
displayed in Figure~\ref{difphases}.

The histogram of the differences between the directions of the
corresponding vectors ($\Delta \varphi$) is presented in
Figure~\ref{histy}. Values of $\Delta\varphi$ slightly anti-correlate
with the average magnitude of the corresponding vectors
($\rho=-0.58$). We think that this is due to uncertainties of both
techniques. From our tests using synthetic data, it became clear
that the inaccuracy of the LCT code is
15~m\,s$^{-1}$ for velocities smaller than 100~m\,s$^{-1}$ and
25~m\,s$^{-1}$ for velocities larger than 100~m\,s$^{-1}$, for both
components \cite{svanda06}. We think that the 10~\% accuracy for the
time-distance velocity vectors is a reasonable estimate. 
As is stated in \citeauthor{zhao01} (\citeyear{zhao01}), cross-talk 
effects between horizontal and vertical components of flow velocities 
affect the time-distance inversion results. The cross-talk 
effect prevents us from inverting the vertical velocity 
correctly, but it does not block the determination of horizontal velocities
(\citeauthor{2003soho...12..417Z}, \citeyear{2003soho...12..417Z}). 
However, the vertical  velocities are not discussed in this paper because they are not 
measured by the LCT technique.
Obviously,
inaccuracy in one component may cause a significant change of the
direction of the horizontal vector for small velocities and, hence,
the agreement of both techniques in such areas is not as good as in
the areas of high velocities.

The vector velocity field may be also influenced by the temporal evolution of 
the traced pattern. We have tested using the full-disc MDI dopplergrams that temporal
changes at mesogranular and smaller scales are effectively filtered out by a $k$--$\omega$
filter. The temporal evolution of the supergranular pattern may, in the worse case, significantly
influence the calculated vector velocity field in the close (roughly equal to the FWHM of the
chosen correlation window) vicinity of a rapidly changing (\emph{e.g.} disappearing) supergranule.

\section{Conclusions}

Flow velocity fields on the solar surface obtained by two different
techniques, time-distance helioseismology and local correlation
tracking (LCT), were compared. Despite the fact that the first
technique uses $p$~modes of solar oscillations to compute the
velocity field (in the data, large-scale structures like
supergranulation are suppressed), while the other one uses
large-scale supergranulation pattern from averaged Doppler images as
tracers for the velocity vectors determination (and requires
$p$~modes removal), we found that both results match reasonably
well. We have confirmed some recent studies that the LCT method
slightly underestimates the actual velocities (as the consequence of
a smoothing procedure), and determined empirical correction factors.
After the corrections, the match in global velocity structures, mean
zonal and meridional flows, is very good. The results of a detailed
comparison of the vector velocity fields are not so satisfactory.
However, the correlation coefficient for individual components of
the flows is positive and significant, so we conclude that a
meaningful match is found. It is shown that the largest disagreement
is caused by very small velocities in some regions, where the errors
of both methods become quite significant.

\acknowledgements M. \v S. has been supported by ESA-PECS under
grant No. 8030 and by the Czech Science Foundation under grant
205/04/2129. M. \v S. would like also thank to the members of the
solar group of Stanford University for all the support during his
stay at Stanford and to his Ph.~D. thesis advisers, Mirek
Klva\v na and Michal Sobotka, for valuable help when tuning the
LCT method. The MDI data were kindly provided by the SOHO/MDI
consortium. SOHO is the project of  international cooperation
between ESA and NASA. We also thank Marc~L.~DeRosa, whose comments 
helped significantly to improve the paper.

\end{article}
\end{document}